# Frequency tripled 1.5 µm telecom laser diode stabilized to iodine hyperfine line in the $10^{-15}$ range

Charles Philippe[1], Rodolphe Le Tragat[1], David Holleville[1], Michel Lours[1], Tuam Minh-Pham[2],
Jan Hrabina[2], Frederic Du Burck[3], Peter Wolf[1] and Ouali Acef[1]
Email: ouali.acef@obspm.fr

[1]LNE-SYRTE, Observatoire de Paris, PSL Research Univ., CNRS, Sorbonne Univ., UPMC Univ. Paris 06, 75014 Paris, France
[2]Institute of Scientific Instruments, Czech Academy of Sciences, Brno, Czech Republic
[3]Laboratoire de Physique des Lasers, Université Paris 13, Sorbonne Paris Cité, 93430 Villetaneuse, France

***Abstract*—** We report on telecom laser frequency stabilization to narrow iodine hyperfine line in the green range of the optical domain, after a frequency tripling process using two nonlinear PPLN crystals. We have generated up to 300 mW optical power in the green ($P_{3\omega}$), from 800 mW of infrared power ($P_\omega$). This result corresponds to an optical conversion efficiency $\eta = P_{3\omega}/P_\omega \sim$ 36 %. To our knowledge, this is the best value ever demonstrated for a CW frequency tripling process. We have used a narrow linewidth iodine hyperfine line (component $a_1$ of the $^{127}I_2$ R 35 (44-0) line) to stabilize the IR laser yielding to frequency stability of $4.8 \times 10^{-14}\tau^{-1/2}$ with a minimum of $6 \times 10^{-15}$ reached after 50 s of integration time. The whole optical setup is very compact and mostly optically fibered. This approach opens the way for efficient and elegant architecture development for space applications as one of several potential uses.

## I. Introduction

Coherent and powerful continuous wave (CW) lasers in the visible are widely needed for various applications such as laser cooling, medicine diagnostics, underwater optical communications, high resolution spectroscopy, etc. [1-5]. On the other hand, IR lasers are currently proposed for many space applications such as gravitational wave detection, inter-satellites or ground to space optical communications, earth observations, etc. [6-8]. These lasers are successfully frequency stabilized against iodine transitions in the green [8-9]. Usually, the optical setups devoted to these space applications are developed in free space configurations.

In this paper, we describe a new approach combining the use of frequency tripled Telecom laser sources (TLS) emitting in the 1.5 µm range and one from thousands of narrow iodine hyperfine lines existing in the green part of the visible spectrum for the frequency stabilization purpose. Compared to other IR laser sources, the TLS exhibit unprecedented intrinsic phase noise (linewidth ~ kHz) associated to extremely small volume (~ cm³) and optically fibered mode operation. On the other hand, iodine lines in the green range of visible domain have remarkable quality factor ($> 2 \times 10^9$) [10], achievable with simple and compact experimental interrogation configurations.

## II. Efficient Frequency Tripling Process

Third harmonic generation (THG) of CW infrared lasers has been demonstrated in only few cases, with very poor efficiency $P_{3\omega}/P_\omega$ [11, 12]. In early 2002, a first attempt to observe iodine lines via a THG of a telecom laser has been described using two second order nonlinear processes in a unique crystal [13]. The two processes were operated in a single periodically poled Lithium Niobate crystal (PPLN) allowing a green power generation at level of few tens of nW. The associated optical conversion from IR to green $\eta = P_{3\omega}/P_\omega$ is in the range of $10^{-5}$ %. The main limitation is due to the difficulty to fulfill the quasi phase matching conditions for SHG and SFG simultaneously in the same nonlinear crystal. Recently, higher efficiency has been demonstrated at level of $\eta = P_{3\omega}/P_\omega = 0.25$ %, corresponding to 1.5 mW generated green light, with two independent crystals used to fulfill two cascaded second order steps: ($\omega + \omega \rightarrow 2\omega$) followed by ($\omega + 2\omega \rightarrow 3\omega$) [14].

In this paper we propose a new optical architecture using two fibered waveguide PPLN nonlinear crystals described in more details in ref. [15]. We utilize two optically fibered waveguide Zn doped PPLN crystals to achieve a second harmonic process (SHG) followed by a sum frequency generation (SFG) in an original optical arrangement as depicted in Fig. 1. We generate up to $P_{3\omega} = 300$ mW at 514 nm using $P_\omega = 200$ mW associated to red power $P_{2\omega} = 330$ mW achieved independently from PPLN1 with 600 mW at $\omega$. Consequently, this maximum output green power was obtained from 0.8 W total IR power at 1.5 µm, corresponding to an optical conversion efficiency $\eta = P_{3\omega} / P_\omega \sim$ 36 %. During measurements reported in Fig. 2, the total optical power incident onto the SFG crystal ($P_\omega + P_{2\omega}$) was intentionally limited to ~ 0.53 W in order to avoid possible optical damage. This new setup including the laser diode, the EDFA and all needed optical components occupy a total volume of 4.5 liters. The IR source used in this work is a butterfly narrow linewidth laser diode (linewidth ~ 2 kHz, power ~10 mW) followed by an erbium doped optical amplifier (EDFA) delivering up to 1 W over the full C band of the telecom range. All optical fibers, splitters and optical isolators used in this setup are polarization maintaining devices. Two homemade electronic control devices are used to fulfill the phase matching temperature conditions within 5 mK.

This work is supported by: Agence Nationale de la Recherche (ASTRID program ANR 11 ASTR 001 01), Labex FIRST-TF, Délégation Générale de l'Armement (DGA), SATT Lutech and AS-GRAM (CNRS/INSU).
C. Philippe PhD thesis is co-funded by Centre National d'Etudes Spatiales (CNES) and SODERN.

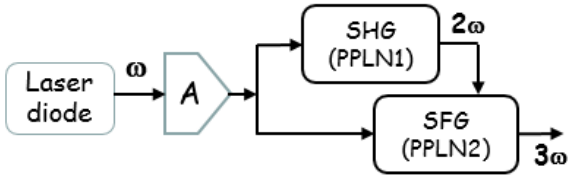

Fig. 1. Principle of the THG setup using two cascaded PPLN crystals: $(\omega + \omega \rightarrow 2\omega) + (\omega + 2\omega \rightarrow 3\omega)$. A= Erbium Doped Optical Fibre Amplifier, PPLN1 = Optical fibered PPLN crystal for SHG operation, PPLN2=optical fibered PPLN crystal for SFG operation.

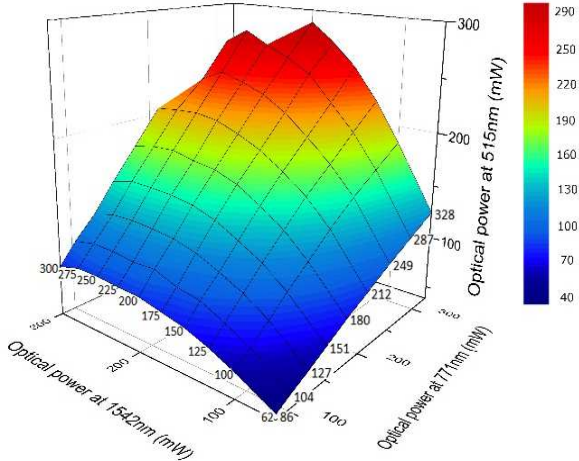

Fig. 2. Evolution of the generated green power (514 nm) versus variations of the red (771 nm) and the IR (1542 nm) powers.

### III. REQUENCY STABILIZATION SETUP

We use the well-known saturated absorption approach associated to the modulation transfer technique to frequency stabilize the 1.542 µm laser diode against iodine hyperfine line in the green (Fig. 3). The pump beam (respectively probe beam) is frequency shifted by 79 MHz (resp. 80 MHz) with acousto-optic modulator. A low frequency modulation (220 kHz) is applied to an electro-optic modulator to detect the atomic saturation signal.

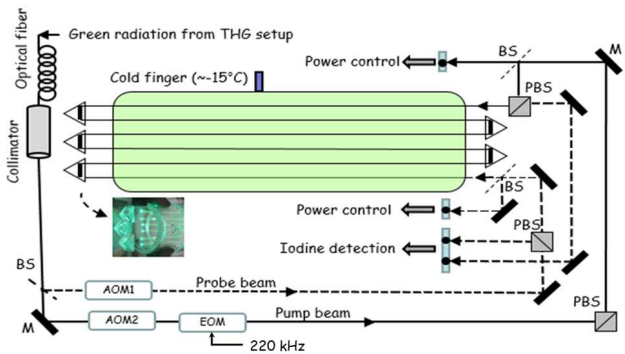

Fig. 3. Iodine stabilization optical setup. BS: beam splitter, M: Mirror, AOM: Acousto-optic modulator, EOM: electro-optic modulator, PBS: polarizing beam splitter.

The two optical laser beams of diameter ~2 mm are carefully collimated and overlapped in the 20 cm long iodine cell. The interaction length is extended up to 1.2 m thanks to 6 successive passes. Both internal and external faces of the cell windows are antireflection coated in the green [16]. The quartz cell is filled with highly pure iodine in Institute of Scientific Instruments in Czech Republic. The saturated absorption signal is detected by a balanced silicon photodiode. A part of the probe beam is split off before propagating in the cell in order to eliminate common noise of the laser probe beam. The probe and pump powers are stabilized with signals detected with two independent photodiodes. An additional photodiode (not shown in fig. 3) is used for a permanent control of the residual amplitude modulation (RAM) associated to the phase modulation of the pump beam. The cold finger temperature of the cell is regulated around -15 °C within 1.5 mK as presented in Fig. 4, using a homemade electronic PID controller. The corresponding vapor pressure in the cell is estimated about 1 Pa.

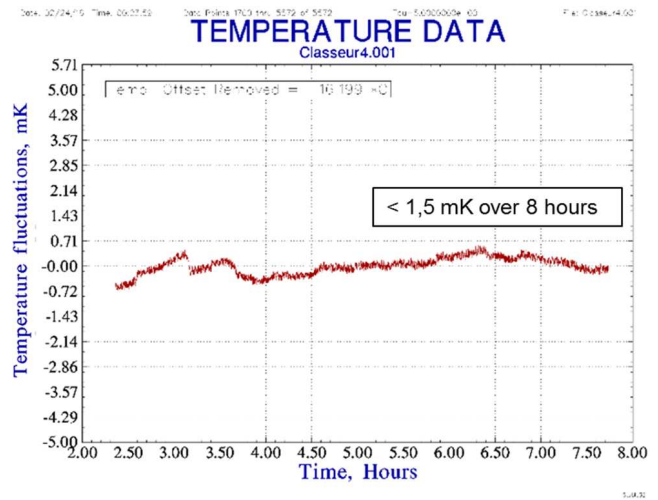

Fig. 4. Evaluation of the cold finger temperature fluctuation

### IV. FREQUENCY STABILITY MEASUREMENT

An independent stable frequency reference laser (FRL) is utilized to fulfill the frequency stability measurement of our 1.5 µm / iodine optical frequency standard. The FRL is based on another IR laser source frequency locked to an ultra-stable optical cavity described elsewhere [17]. It is located in a separate building and is connected to our experiment by a 200 meters optical fiber link, as seen in Fig. 5. During this preliminary measurement the frequency noise of this optical link was not compensated, because its contribution together with the reference cavity instability exhibit an Allan deviation at level of ~$10^{-15}$ over the full integration time measurement. Subsequently, the frequency stability evaluation of our iodine stabilized laser is not affected. The beat note between our iodine frequency stabilized laser and the reference laser exhibits a linewidth smaller than 5 kHz (Fig. 5).

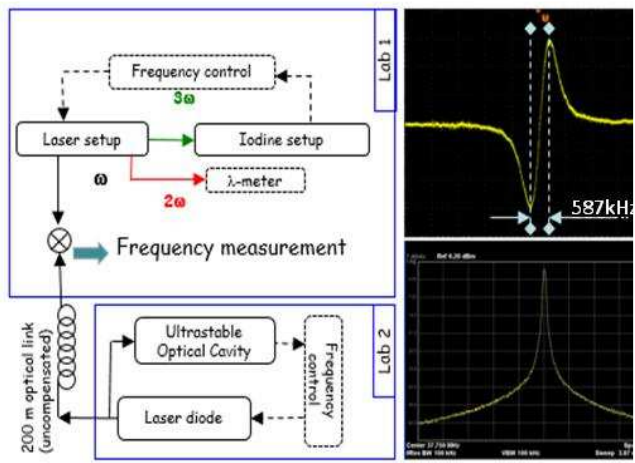

Fig. 5. Optical setup arrangement developed for the frequency stabilization measurement. Top right figure shows the line utilized to stabilize the laser. It corresponds to the $a_1$ hyperfine component of the R 35 (44-0) $^{127}I_2$ line at 514.017 nm [18]. The figure bottom right reports the beat note in the infrared between the reference and the stabilized lasers (width ~ 5 kHz).

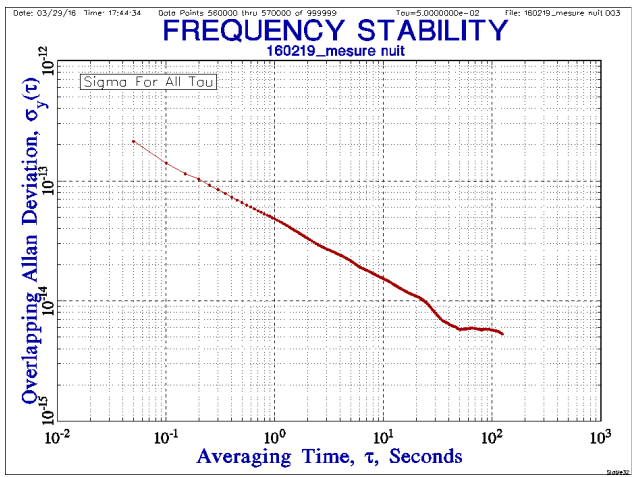

Fig. 6. Evaluation of the cold finger temperature fluctuation

Fig. 6 shows our preliminary best result with an Allan deviation decreasing with a slope of 4.8 x $10^{-14}$ $\tau^{-1/2}$ with a minimum value of 6 x $10^{-15}$ for 50 s of integration time. During this preliminary frequency stability measurement the residual amplitude modulation (RAM) was not compensated and could explain the behavior of the frequency stability observed for $\tau > 50$ s.

V. CONCLUSION

We have developed a compact device for the generation of an intense continuous radiation (300 mW) in the green from 800 mW of IR. We have shown the possibility of using the narrow transitions of molecular iodine absorption spectrum to stabilize the laser diode frequency.

Frequency stability in the range of $10^{-15}$ was demonstrated using a small part of the green power (< 10 mW).